# Low Latency Techniques for Mobile Backhaul over DOCSIS

John T. Chapman[1], Jennifer Andreoli-Fang[2], Michel Chauvin[2], Elias Chavarria Reyes[1],
Zheng Lu[1], Dantong Liu[1], Joey Padden[2], Alon Bernstein[1]

[1]Chief Technology and Architecture Office, Cisco, San Jose, CA
[2]Office of the CTO and R&D, CableLabs, Boulder, CO

*Abstract*—The mobile network operators (MNOs) are looking into economically viable backhaul solutions as alternatives to fiber, specifically the hybrid fiber coaxial networks (HFC). When the latencies from both the wireless and the HFC networks are added together, the result is a noticeable end-to-end system latency, particularly under network congestion. In order to decrease total system latency, we proposed a method to improve upstream user-to-mobile core latency by coordinating the LTE and HFC scheduling in previous papers. In this paper, we implement and optimize the proposed method on a custom LTE and DOCSIS end-to-end system testbed. The testbed uses the OpenAirInterface (OAI) platform for the LTE network, along with Cisco's broadband router cBR-8 that is currently deployed in the HFC networks around the world. Our results show a backhaul latency improvement under all traffic load conditions.

## I. INTRODUCTION

One of the next big opportunities for cable operators is the mobile market. For a long time, cable and mobile operators have been competitors, both competing for voice or Internet subscribers. The mobile network operators (MNOs) under pressure to deploy dense small cell networks in response to the tremendous growth in mobile data usage [1]. Consequently, the MNOs are increasingly looking at cable networks as a backhaul solution for small cells. Fiber has been traditionally preferred by the MNOs to backhaul macro cells. However, fiber is sparse, and to install fiber ubiquitously is not as economical as using existing fixed infrastructure. Cable operators have built and deployed HFC networks everywhere to service broadband residential and commercial customers. This HFC network is an attractive option for small cell backhaul as it is ubiquitous, has ample capacity, and can provide a lower cost alternative to running new fiber.

Today's HFC network incurs higher latency than the allocated timing budget for mobile backhaul. The Data Over Cable Service Interface Specifications (DOCSIS®) [2][3] defines the broadband communication protocol between the cable modems (CMs) and the cable modem termination system (CMTS) over the HFC network. The DOCSIS downstream incurs about 1 ms of latency. It accounts for a fraction of the round trip latency. Hence, we focus on improving the DOCSIS upstream, which today incurs a minimum latency of 5 ms, average of 11-15 ms, and can be 20 to 50 ms for a loaded network [4]. In comparison, fiber networks have been measured to incur 2-3 ms of latency [9]. The LTE backhaul budget is lower than the minimum DOCSIS latency [5]. Interference coordination techniques such as Coordinated Multi-Point (CoMP) require 5 ms of total X2 latency to realize significant gain [6][7]. Furthermore, most of 5G applications require less than 10 ms user-to-core latency, with some ultra-low latency applications requiring 1 ms latency [8][10]. These requirements are difficult to meet with DOCSIS today.

Both LTE and HFC networks follow a request-grant-data (REQ-GNT) loop for typical upstream transmissions. When wireless data is sent from the user equipment (UE) to the LTE base station (eNB) and backhauled over DOCSIS, the data experiences this type of 3-way loop twice. This occurs when the two networks, the wireless and its backhaul, are uncoordinated.

In previous papers [16][17] we proposed a Bandwidth Report (BWR) based method to improve the upstream UE-to-mobile core latency by coordinating the LTE and DOCSIS scheduling operations. Previous simulation results showed the latency reduction potential of the method.

In this paper, we present the implementation of the BWR algorithm on a custom, real-time LTE-DOCSIS testbed using the OpenAirInterface (OAI) LTE platform, along with an application programming interface (API) built into Cisco's commercially deployed broadband router cBR-8 on the DOCSIS link. Optimization is made on calculating the expected egress times for the LTE data, such that just-in-time DOCSIS grants are ensured. A key benefit of the BWR method is that the underlying DOCSIS scheduler remains unchanged. Instead, an API is inserted before the DOCSIS scheduler to receive and interpret the LTE information, and convert it into traditional DOCSIS bandwidth requests that are inputs to the existing DOCSIS scheduler. This reduces the amount of software engineering, which enabled us to rapidly prototype the system and in turn will enable rapid industry adoption.

This paper is organized as follows. Section 2 provides a brief introduction on the DOCSIS and LTE upstream scheduling operations. Section 3 discusses the proposed method in detail. Section 4 describes the real-time testbed in detail and shows latency results.

## II. BACKGROUND AND PREVIOUS WORK

### A. DOCSIS Upstream Data Plane Latency

The DOCSIS upstream uses a REQ-GNT approach as shown in Fig. 1a. A request (REQ) message is sent from the CM to the CMTS. The CMTS prepares a DOCSIS control frame, called the MAP, and inserts an entry indicating a grant for the CM. The grant entry in the MAP message contains an upstream service identifier (SID) associated with a service flow assigned to the CM, a transmission time signaled as a slot number, and the number of bytes to be transmitted. The CMTS transmits the MAP to the CM, so that the CM can make use of the grant to send its upstream data to the CMTS.

Various factors determine the request-grant time in DOCSIS. The DOCSIS specification contains a detailed analysis of the REQ-GNT delay [2]. The minimum delay is every second MAP time plus some circuit and queuing delay. The DOCSIS 3.0 CMTSs currently use a 2 ms MAP time. One published set of test results for a Cisco uBR10K CMTS showed a 5 ms minimum REQ-GNT time [4].

The average latency is expected to be higher due to the random arrivals of data and therefore, REQs, as well as the scheduling of contention regions. Higher network loads result in a higher probability of collisions among the REQs, which trigger a truncated exponential backoff that further increases latency. The average DOCSIS upstream latency for best effort traffic is measured [4] to be 11-15 ms. The maximum upstream latency can be significantly higher.

The CMTS scheduling algorithm can have a significant impact on REQ-GNT delay. If the CMTS uses the best effort (BE) scheduling algorithm, REQs are sent in contention slots and could collide. This may result in higher latency than the real-time polling service (RTPS) scheduling algorithm where the REQs are placed in dedicated slots that ensures successful transmission. When upstream load is low, BE can actually perform better than RTPS. This is because most of upstream resources can be used as contention request slots. REQs can thus be sent immediately rather waiting to be sent in dedicated slots. Therefore, in the low loading case, BE results in lower latency than RTPS. However, in a busy system where more upstream resources are used for data transmission, the CMTS schedules fewer contention request slots. The CMs have to wait longer to send REQs and retransmit REQs when they collide. In this case, RTPS can provide guaranteed latency.

### B. LTE Uplink Data Plane Latency

At the heart of the LTE uplink scheduler is a similar REQ-GNT mechanism, shown in Fig. 1b. The LTE request is sent from the UE to an eNB. When data arrives at the UE, the UE first determines if it already has an LTE uplink (UL) grant. If it does not, the UE waits for an opportunity to transmit a scheduling request (SR), where typical SR opportunities can come along once every 1 to 10 ms depending on the configured periodicity of the SR. The purpose of the SR is to keep the upstream signaling overhead low, which is critical in wireless where spectrum is costly.

Upon receiving the SR, the eNB schedules an UL grant so that the UE can transmit a buffer status report (BSR) to the eNB. Once the eNB receives the BSR from the UE, it schedules UL grants and transmits the grants to the UE. Due to processing constraints, the eNB-UE and UE-eNB delays are typically 4 ms. In other words, upon receiving the BSR, the eNB performs scheduling and sends Downlink Control Information 0 (DCI0) to arrive at the UE in 4 subframes, and the UE is scheduled to transmit 4 subframes later. Assuming an SR periodicity of 5 ms, the minimum delay before a UE can transmit UL data is 18 ms which is longer than the minimum DOCSIS request-grant latency. Detailed request-grant latency calculations are shown in Table I.

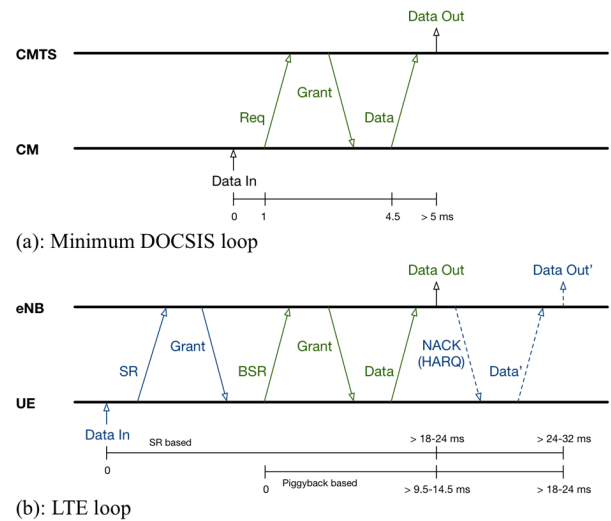

Fig. 1. DOCSIS and LTE REG-GNT-data loops

TABLE I      LTE UPLINK LATENCY COMPONENTS

| Latency components | Latency (ms) |
|---|---|
| Waiting time for SR (assume configured SR period of 5 ms) | 0.5 – 5.5 |
| UE sends SR, eNB decodes SR, eNB generates grant for BSR | 4 |
| eNB sends grant, UE processes grant, UE generates BSR | 4 |
| eNB processes BSR, eNB generates grant for data | 4 |
| eNB sends grant, UE processes grant, UE sends UL data | 4 |
| eNB decodes UL data (estimate) | 1.5 – 2.5 |
| Total | 18 – 24 |

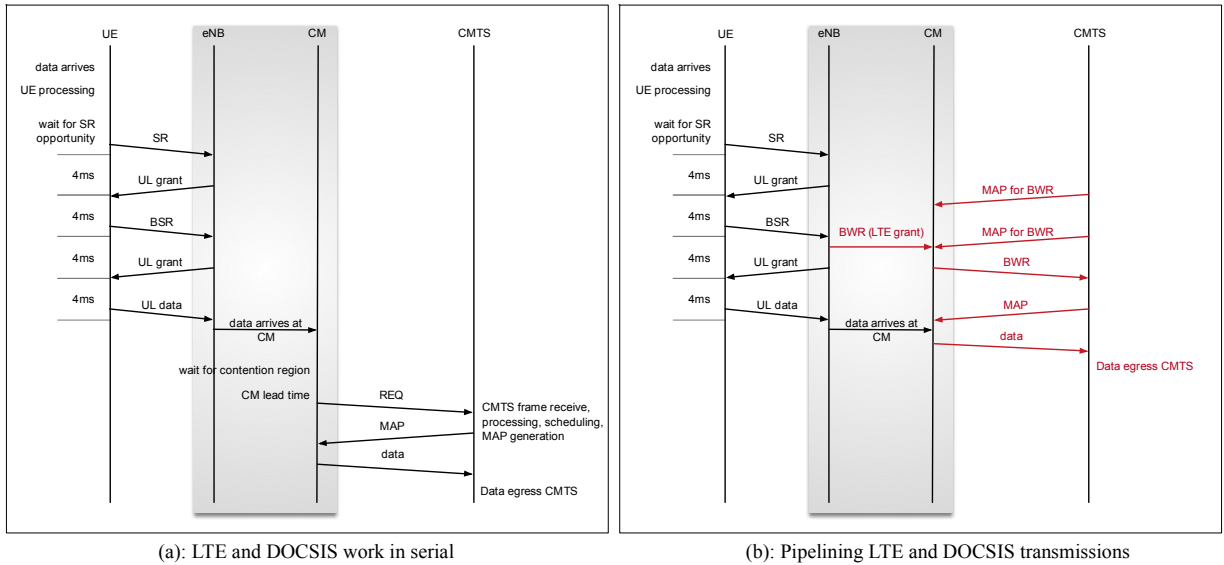

Fig. 2. Backhauling LTE Data with a DOCSIS Network

*C. Sum LTE and DOCSIS Latencies Altogether*

Referring to Fig. 2a, when uplink data destined for the mobile core arrives at the UE, it needs to traverse two 3-way REQ-GNT-data loops in serial. The first REQ-GNT loop is started by the UE using either an SR or a BSR, depending on if the UE is already scheduled. The eNB then grants the UE UL resources that allows the UE to transmit data to the eNB. The eNB reassembles the UL data and transmits complete packets to the CM, where a DOCSIS REQ will be transmitted by the CM to the CMTS. The CMTS then assigns DOCSIS grant(s) that allow the data to traverse through the DOCSIS backhaul which it eventually reaches the mobile core. The LTE and DOCSIS latencies are added in serial.

*D. Previous Work*

To the best of our knowledge, there has been no previous attempt to reduce latency by coordinating wireless and backhaul networks. Efforts have been made to separately reduce the LTE latency and the DOCSIS latency. The 3GPP is working on techniques including contention based data transmissions, and reducing the transmission time intervals [11]. However, the proposed techniques have yet to be standardized into the 3GPP Stage 2 specifications.

DOCSIS has two built-in mechanisms to address latency: Unsolicited Grant Service (UGS)/RTPS and Active Queue Management (AQM). The CMTS uses UGS scheduling to eliminate the REG-GNT latency by periodically scheduling fixed sized grants. However, UGS cannot be used to backhaul LTE traffic, as UGS grants are fixed in size, and packets cannot be fragmented. RTPS eliminates the contention latency by allowing the REQs to be sent in dedicated slots rather than in contention slots as in BE. While RTPS eliminates contention latency, it does not address the latency incurred due to the REQ-GNT loop. The AQM [12] feature does not address the scheduling latency, as it only applies when multiple applications share a network connection and requires changes to the modem hardware. Besides the standardized mechanisms, the CMTS vendor Cisco has implemented a predictive MAC scheduler, which proactively grants in an intelligent manner [4].

The work described in this paper differ from all these prior efforts, but can operate closely with a predictive scheduler, as it enables the scheduler to more accurately predict the future upstream data flow across the DOCSIS backhaul. Our latency reduction method can be implemented in an API outside the CMTS scheduler to avoid changing a critical section of the CMTS code. This new CMTS API accepts the information about the expected future LTE UL transmission, and provides this information to the DOCSIS scheduler. Because of its simplicity, our latency reduction method works with existing CMTS schedulers and requires minimal software engineering, thereby facilitating significantly faster industry adoption.

## III. PIPELINING THE LTE AND DOCSIS SCHEDULING OPERATIONS

This section describes a method to improve the upstream latency by coordinating the LTE and DOCSIS scheduling loops. Since the DOCSIS REQ-GNT-data loop is the main contributor of backhaul upstream latency, the REQ-GNT-data processes on LTE and DOCSIS links are pipelined to reduce latency.

The first stage of the pipeline consists of the LTE REQ-GNT-data operation. At the time a UL scheduling decision is made by the eNB, the eNB also indicates to the CM that it is expecting UL data in the near future. This information prompts the second stage of the pipeline, the DOCSIS 3-way loop, to start early. Given the LTE transmission information, the DOCSIS scheduler prepares a grant to arrive at the CM just in time to transport the LTE data when it arrives at the CM. Using our method, the upstream operation through the two networks becomes pipelined, where the first pipeline stage is the LTE scheduler, which informs the next stage, the DOCSIS scheduler, of what is about to come.

The details of the pipelining operations are shown in Fig. 2b. After the eNB has received the BSR and generated an LTE UL grant, the eNB MAC generates a Bandwidth Report (BWR) message, indicating the size of the grant and the time when the UL data is expected to arrive at the CM. This message is then received by the CM and forwarded to the CMTS transparently. Generally, the eNB and the CM are connected via a Gigabit Ethernet connection and the propagation time of the message is negligible.

In order to ensure the BWR messages arrive at the CMTS on time, the CMTS periodically grants the CM via low latency service flow such as RTPS or UGS to transport the BWRs. Since the eNB scheduler generates UL data grants every subframe, the optimal BWR generation and DOCSIS grant interval is 1 ms, but the method also works well with slightly longer grant intervals such as 2 or 4 ms. The CMTS scheduler uses the information included in the BWR message to schedule a transmission time for the UE data, and generates a US grant allocation MAP indicating the scheduled transmission start time to the CM. When LTE UL data egresses the eNB and arrives at the CM, a DOCSIS upstream (US) grant has arrived and can then be used immediately by the CM to forward the LTE data to the CMTS.

### A. The Bandwidth Report

The BWR needs to be sent to the CMTS quickly in order for the CMTS scheduler to make use of the time sensitive information and to generate an US grant. One way to achieve this goal is to transport the BWR message via a RTPS or a UGS service flow, since it allows the CM to skip the contention request step when sending the BWR.

In LTE, multiple UEs can be scheduled at the same time, and each UE's UL data is reported based on priority. There are two options to describe the UL traffic to the CMTS scheduler: aggregate all the grants of all the UEs, or separate the grants based on priority.

In the first option, the BWR can simply describe the LTE grant as a "bulk grant" in terms of a single block of bytes that the eNB scheduler has allocated for transmission in future subframes. Since multiple UEs may be scheduled for transmission in a single UL subframe, the eNB MAC simply aggregates the individual LTE grants allocated for each UE in that subframe in a single block of bytes.

If traffic differentiation is desired on the DOCSIS link, the BWR can report the amount of UL grant for each Logical Channel Group (LCG). The eNB MAC includes 4 blocks of bytes in the BWR, with each block indicating the aggregate number of bytes allocated for an LCG for all the UEs scheduled to transmit in the future subframe described by the BWR. Because each LCG maps to a different set of QoS parameters, the CMTS scheduler uses this information to prioritize the LTE data traffic in a more granular manner.

The BWR takes up a fraction of the DOCSIS upstream capacity. The optimal period for BWR for LTE FDD is 1 ms because the eNB scheduler makes scheduling decision once every 1 ms. For example, an 80-byte BWR transmitted every 1 ms consumes 640 kbps per eNB. This consumption is not affected by the LTE UL data traffic load.

We note that LTE uses Hybrid Automatic Repeat reQuest (HARQ) to increase reliability and reduce latency associated with air interface transmission failures. HARQ failures will translate into the under-utilization of the DOCSIS just-in-time grants requested by the BWR. The topic has been discussed in detail in [17] and will not be repeated here.

### IV. REAL-TIME EXPERIMENTATION

In this section, we describe the real-time system testbed that was built with open source and commercially available equipment and present latency improvement results from the testbed.

### A. Testbed Description

The OpenAirInterface (OAI) [18] platform was chosen to prototype the BWR in the LTE portion of the system. The OAI platform is an open source solution that implements a 3GPP LTE Rel-8 full stack eNB in software and allows for flexible experimentation with real LTE networks. It is fully integrated with off-the-shelf software defined radios (SDR), and supports communication with a host of EPCs. For our testbed, we used LimeSDR and the OAI EPC.

The DOCSIS portion of the network uses a commercially available ARRIS DOCSIS 3.1 cable modem and a Cisco broadband router cBR-8 that serves as the CMTS. The cBR-8 is currently deployed by cable operators around world. The cBR-8 and the OAI stack are synchronized using an Adva OSA 5420 PTP Grandmaster clock which communicates using the IEEE 1588 Precision Time Protocol (PTP).

The BWR method does not require engineering new software into the cable modem. Code was added to the OAI eNB to generate BWR messages. An API was implemented on the cBR-8 to receive and interpret the BWR messages, and to translate them into a DOCSIS REQ as an input into the existing DOCSIS scheduler on the cBR-8. As such, no software modification is needed for the existing DOCSIS MAC scheduler.

The testbed architecture is shown in Fig. 3, while a portion of the physical testbed is shown in Fig. 4. The cBR-8, the EPC, and the Adva Grandmaster are not shown because they are remotely connected. For latency measurements on a link, a ntop real-time packet capture software running on commodity hardware with hardware timestamping capability using Intel i350 NIC is placed at various interfaces to record the relative timestamps of the packets traversing the network.

TABLE II shows the experimental setup parameters. To simplify the amount of information that has to be processed, we configured the backhaul to use only 1 DOCSIS upstream channel, with the channel width to be 1.6 MHz at 1.28 Msymbol per second. In deployments, the DOCSIS systems are capable of much wider channel width with up to 96 MHz on the upstream. With channel bonding, DOCSIS can achieve multi-gigabits per second speeds. Our simplified configuration does not impact test results, as our experiments use low bandwidth ping packets to test the latency.

To achieve optimal latency gain, a BWR message should be generated every LTE subframe. At the time of writing, the production code on the cBR-8 generates a MAP every 2 ms so we aggregate the BWR requested data size over 2 LTE subframes to match the MAP interval and UGS grant interval in our experiments. In addition, although DOCSIS does not restrict UGS grant interval, the minimum UGS grant interval supported by the production code on the cBR-8 is 4 ms at the time of writing. Therefore, in the testbed we configured the eNB to send 2 BWRs every UGS grant interval. The CMTS API that receives the BWR should be capable of using the timestamp inside the BWR to allocate just-in-time DOCSIS grants for transporting the ping packets on the DOCSIS link as soon as they arrive at the CM queue.

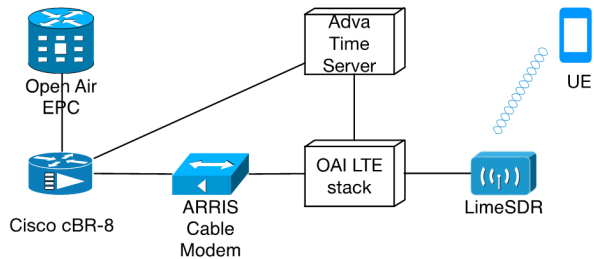

Fig. 3. Testbed architecture.

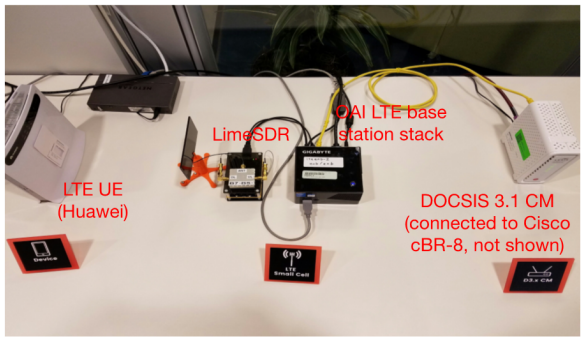

Fig. 4. Testbed setup

For the experiments, we set up the testbed to generate different amount of background traffic to reflect varying degrees of network operating conditions. In order to provide background traffic, the CM under test (CM-UT) shares the DOCSIS upstream channel with 10 other background CMs (CM-BGs). The CM-UT provides backhaul service for a BWR-capable eNB, while the CM-BGs are simply background traffic generators. Each CM-BG is configured with 3 DOCSIS upstream service flows. The total DOCSIS upstream traffic generated by the 10 CM-BGs is selected to produce various network loading points for the purposes of evaluating the performance increase due to the BWR.

For each network load point, we ran end-to-end ping tests from the UE to the EPC with background loading to show the potential of latency reduction using the BWR method at varying load points.

TABLE II  TESTBED SETTINGS

| DOCSIS parameters | |
|---|---|
| Number of CMs | 1 device under test (CM-UT) <br><br> 10 background devices (CM-BG) when background traffic is used |
| Channel configurations | 4 downstream, 1 upstream |
| Upstream PHY | Channel bandwidth = 1.6 MHz <br> Symbol rate = 1.28 MSym/sec <br> Modulation = 64 QAM <br> Raw PHY throughput = 7.68 Mbps |
| Scheduling service for BWR | UGS <br> Grant interval = 4ms <br> Grants per interval = 2 <br> Grant interval jitter = 500 μs <br> Grant size: 90 bytes |
| Scheduling service for data | Best effort, for both CM-UT and CM-BGs |
| **LTE parameters** | |
| LTE channel bandwidth | 5 MHz (25 PRBs) |
| Duplexing method | FDD |
| Spatial multiplexing | SISO |
| HARQ | OFF or ON (10 % BLER) |
| Number of UE | 1 |
| BWR tx periodicity | 2 ms |

For each load point, we first consider the baseline performance where no coordination exists between the DOCSIS and LTE networks. This is achieved by disabling the BWRs on the eNB, and disabling the API that receives and processes the BWRs on the CMTS. We then switch on the BWR generation at the eNB and the API on the CMTS and study the proposed BWR method where the eNB signals the expected future LTE transmission information to the CMTS using the BWR and the CMTS issues just-in-time DOCSIS grants for the LTE upstream data. A laptop is connected to the UE to send ping packets. The latency is measured by the ntop sniffer with hardware timestamping capability located at various interfaces of the system. In particular, we measured the average ping latency across the LTE air interface link, which ranges from 36-38 ms.

To measure the performance gain of the BWR method over the baseline LTE-DOCSIS network, we sent ping packets at an interval of 20 ms, for each DOCSIS upstream load point. The ping packet payload size was ramped from 64 to 1280 bytes (92 to 1308 bytes with overhead), in steps of 64 bytes to ensure that segmentations occur on the LTE air interface. Each test was run for 10 minutes and repeated 6 times.

*B. Test Results*

Fig. 5 shows the average round trip time (RTT) results for the baseline scenario (black), BWR ON scenario (green) and the latency gain (orange) achieved by the BWR method for a range of DOCSIS network loads between 8% and 70%. As mentioned earlier in this section, the average LTE latency in our system is measured to be 36-38 ms regardless of the DOCSIS load. Because the BWR does not affect LTE air interface latency, the latency gain shown in Fig. 5 occurred entirely on the DOCSIS link and was the result of using the BWR method. The latency reduction for the DOCSIS link as a result of the BWR-based method ranged from 55% when no background traffic was used, to 68% for 70% network load. (Note: percentage increase is computed for DOCSIS link only, i.e., subtracting 38 ms of measured LTE latency from total RTT in Fig. 5.)

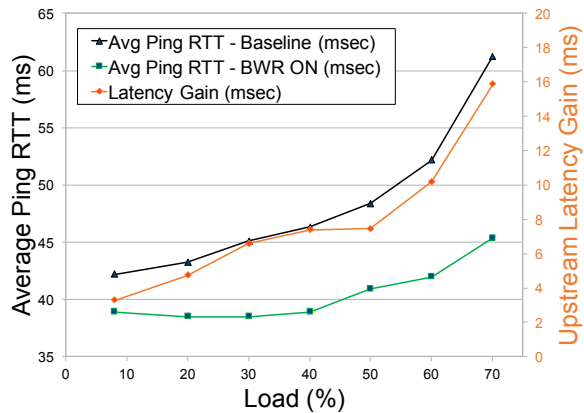

Fig. 5. Average Latency vs. Load

## V. CONCLUSION

In this paper, we described a method to improve the upstream system latency for a mobile network backhauled over a DOCSIS network. We built a real-time LTE-DOCSIS custom testbed using the OAI LTE platform and a widely deployed commercial CMTS. Due to the simplicity of the BWR approach, the BWR was implemented on the testbed without a significant amount of software engineering. This should facilitate rapid industry adoption. Real-time experimentation shows that the BWR algorithm significantly reduces the backhaul latency across a wide range of network operating points.

ACKNOWLEDGMENT

Many engineers provided technical insights throughout this project. The lead authors from CableLabs and Cisco would like to thank Balkan Kecicioglu and Vaibhav Singh of CableLabs, and Oliver Bull of Cisco for the technical discussions.